\documentclass[pra,aps,twocolumn,showpacs]{revtex4}
\usepackage{times,epsfig,amssymb,amsfonts,amsmath}%\usepackage{epsfig}
\usepackage{cases}

\newcommand{\ket}[1]{|#1\rangle}
\newcommand{\bra}[1]{\langle #1|}
\newcommand{\proj}[1]{\ket{#1}\bra{#1}}

\begin{document}

\title{Entanglement and measurement-induced nonlocality of mixed maximally entangled states
in multipartite dynamics}

\author{Li-Die Wang$^{1}$}

\author{Li-Tao Wang$^{1}$}

\author{Mou Yang$^{2}$}

\author{Jing-Zhou Xu$^{1}$}

\author{Z. D. Wang$^{3}$}
\email{zwang@hku.hk}

\author{Yan-Kui Bai$^{1,3}$}
\email{ykbai@semi.ac.cn}

\affiliation{$^1$ College of Physics Science and Information Engineering and Hebei Advanced Thin Films
Laboratory, Hebei Normal University, Shijiazhuang, Hebei 050024, China\\
$^2$ Laboratory of Quantum Engineering and Quantum Materials, School of Physics and Telecommunication
Engineering, South China Normal University, Guangzhou 510006, China\\
$^3$ Department of Physics and Centre of Theoretical and Computational Physics, The University of Hong
Kong, Pokfulam Road, Hong Kong, China}

%%%%%%%%%%%%%%%%%%%%%%%%%%%%%%%%%%%%%%%%%%%%%%%%%%%%%%%%%%%%%%%%%%%%%%%%%%%%%%%%%%%%%%%%%%%%%%%%%
\begin{abstract}
The maximally entangled state can be in a mixed state as well as the well-known pure state. Taking
the negativity as a measure of entanglement, we study the entanglement dynamics of bipartite,
mixed maximally entangled states (MMESs) in multipartite cavity-reservoir systems. It is found that
the MMES can exhibit the phenomenon of entanglement sudden death, which is quite different from
the asymptotic decay of the pure-Bell-state case. We also find that maximal entanglement cannot
guarantee maximal nonlocality and the MMES does not correspond to the state with maximal
measurement-induced nonlocality (MIN). In fact, the value and dynamic behavior of the MIN for the
MMESs are dependent on the mixed state probability. In addition, we investigate the distributions
of negativity and the MIN in a multipartite system, where the two types of correlations have
different monogamous properties.
\end{abstract}
%%%%%%%%%%%%%%%%%%%%%%%%%%%%%%%%%%%%%%%%%%%%%%%%%%%%%%%%%%%%%%%%%%%%%%%%%%%%%%%%%%%%%%%%%%%%%%%%%


\maketitle

\section{Introduction}

The maximally entangled state plays an important role in quantum information processing
\cite{horo09rmp,elts14jpa,cui12nc}, including quantum teleportation \cite{ben93prl}, quantum
cryptographic protocols \cite{ekert91prl} and quantum dense coding \cite{ben92prl}. In bipartite
$d\otimes d$ systems, Cavalcanti \emph{et al} proved that all maximally entangled states are pure
states \cite{cava05pra}. For example, in the simplest $2\otimes 2$ systems, the pure maximally
entangled state is the Bell state which can be written as
\begin{equation}\label{1}
\ket{\psi}=(\ket{00}+\ket{11})/\sqrt{2}
\end{equation}
up to local unitary transformations. Recently, it was further shown that there exist mixed
maximally entangled states (MMESs) in bipartite $d\otimes d{'}$ systems with $d{'}\geq 2d$, which can
be used as a resource for faithful teleportation \cite{zgli12qic,mjzhao15pra}. The MMES in
$2\otimes 4$ systems has the form \cite{zgli12qic}
\begin{equation}\label{2}
\rho=p\proj{\psi_1}+(1-p)\proj{\psi_2},
\end{equation}
where the mixed state probability $p$ lies in the range $(0,1)$, and the two pure state components
are $\ket{\psi_1}=(\ket{00}+\ket{11})/\sqrt{2}$ and $\ket{\psi_2}=(\ket{02}+\ket{13})/\sqrt{2}$.

The dynamic behavior of entanglement is a fundamental property of quantum systems. This is because
unavoidable interactions with the environment may lead the entanglement of quantum systems to be
degraded and, in certain cases, to disappear in a finite time (i.e., the so-called entanglement sudden
death, ESD) \cite{zycz01pra,sche03jmo,tyu04prl,alm07sci,lau07prl,tyu09sci}.
L\'{o}pez \emph{et al} analyzed the dynamic behavior of entangled cavity photons being affected by
two dissipative reservoirs \cite{lopez08prl} and found that the entanglement of cavity photons
initially in the two-qubit Bell state decays in an asymptotic manner. However, for the newly
introduced MMES, its entanglement dynamic property is still an open problem, especially for a real
quantum system. Since the MMES is a perfect physical resource in quantum information processing
\cite{zgli12qic,mjzhao15pra}, it is desirable to investigate its dynamical property in a quantum
system. This is because, once the entanglement evolution experiences the ESD, we are
no longer able to concentrate the entanglement of MMES, which results in some entanglement-based quantum
communication protocols losing their efficacy. In this sense, the study of the entanglement dynamic 
property of the MMES can provide not only useful knowledge for practical quantum operations but also 
the necessary information to cope with the decay of entanglement.

Nonlocality is also a kind of resource in quantum information processing \cite{ben99pra} and has a close
relationship with quantum entanglement \cite{buhr10rmp}. The measurement-induced nonlocality (MIN)
\cite{sluo11prl} is the maximum global effect caused by locally invariant measurement, which is
different from the conventionally mentioned quantum nonlocality related to
the violation of Bell's inequalities \cite{bell64phy,clau69prl}. Moreover, the MIN can quantify the
nonlocal resource in quantum communication protocols involving local measurement and a comparison
between the pre- and postmeasurement states \cite{sluo11prl}. Luo and Fu proved that, for the pure Bell
state, the MIN
achieves the maximal value \cite{sluo11prl}. But it is not clear whether or not the MMES also has the
maximal nonlocality. In particular, can the maximal entanglement guarantee the maximal nonlocality?
Furthermore, in order to obtain a deep understanding of the dynamic properties of MMESs, it is
helpful to analyze the entanglement and nonlocality distributions in an enlarged system
including its environment.

In this paper, as quantified by entanglement negativity \cite{vidal02pra} and the MIN \cite{sluo11prl},
we study the dynamic properties of the MMES in the dissipative procedure of multipartite
cavity-reservoir systems. It is found that the MMES can disentangle in a finite time, which is quite
different from the asymptotical decay of a pure Bell state. We also find that the MIN of the MMES is not
maximal, which means that the maximal entanglement cannot guarantee the maximal nonlocality. In
addition, the evolution of the MIN is dependent on the mixed probability of the MMES. Finally, we
investigate the distributions of the negativity and the MIN in the multipartite system, where the
squared negativity is monogamous but the MIN is not monogamous.

\section{Dynamic properties of Entanglement and nonlocality for the MMES}

We first recall the definition of the MMES before analyzing its dynamic properties. In $d\otimes d'$
systems, a mixed state is an MMES if and only if it has the form \cite{zgli12qic,mjzhao15pra}
\begin{equation}\label{3}
\rho=\sum_{m=1}^{K}p_m\proj{\psi_m},
\end{equation}
where the mixed state probabilities satisfy $\sum_{m=1}^{K} p_m=1$ with
$K\leq \mbox{floor}(d^{\prime}/d)$, and the pure state component is
\begin{equation}\label{4}
\ket{\psi_m}=\frac{1}{\sqrt{d}}\sum_{i=0}^{d-1}\ket{i}\otimes\ket{i+(m-1)d},
\end{equation}
which is the maximally entangled pure state. In the following, we will study the dynamic properties of
MMESs in bipartite $2\otimes 4$ systems.

We consider a practical dynamic system of two cavities interacting with two independent
reservoirs. The initial state of the four-partite system is
\begin{equation}\label{5}
\rho_{c_1c_2r_1r_2}(0)=\rho_{c_1c_2}(0)\otimes \proj{00}_{r_1r_2}
\end{equation}
where the two reservoirs are in the vacuum state, and the two cavities are in an MMES,
\begin{equation}\label{6}
\rho_{c_1c_2}(0)=p\proj{\psi_1}+(1-p)\proj{\psi_2}
\end{equation}
with $\ket{\psi_1}=(\ket{00}+\ket{11})/\sqrt{2}$ and $\ket{\psi_2}=(\ket{02}+\ket{13})/\sqrt{2}$.
It should be noted that, although $\rho_{c_1c_2}$ is written as a probability mix of $\ket{\psi_1}$
and $\ket{\psi_2}$, its pure-state component has the generic form $\sqrt{q}\ket{\psi_1}+e^{i\phi}
\sqrt{1-q}\ket{\psi_2}$, with the parameters $q\in[0,1]$ and $\phi\in[0,2\pi]$.
The interaction of a single cavity and an $N$-mode reservoir is described by the
Hamiltonian \cite{lopez08prl,byw09pra,wen11epjd,ali14pra}
\begin{equation}\label{7}
\hat{H}=\hbar \omega \hat{a}^{\dagger}\hat{a}+\hbar\sum_{k=1}^{N}\omega_{k}
\hat{b}_k^{\dagger}\hat{b}_k+\hbar\sum_{k=1}^{N}g_{k}(\hat{a}
\hat{b}_{k}^{\dagger}+\hat{b}_{k}\hat{a}^{\dagger}).
\end{equation}
At later times, in the limit $N\rightarrow\infty$ for the reservoirs with a flat spectrum
\cite{lopez08prl}, the state is given by
\begin{eqnarray}\label{8}
\rho_{c_1r_1c_2r_2}(t)
&=&\frac{p}{2}[(\ket{\phi_0}_{c_1r_1}\ket{\phi_0}_{c_2r_2}
+\ket{\phi_1^t}_{c_1r_1}\ket{\phi_1^t}_{c_2r_2})\nonumber\\
&&\times (\bra{\phi_0}_{c_1r_1}\bra{\phi_0}_{c_2r_2}
+\bra{\phi_1^t}_{c_1r_1}\bra{\phi_1^t}_{c_2r_2})]\nonumber\\
&+&\frac{1-p}{2}[(\ket{\phi_0}_{c_1r_1}\ket{\phi_2^t}_{c_2r_2}
+\ket{\phi_1^t}_{c_1r_1}\ket{\phi_3^t}_{c_2r_2})\nonumber\\
&&\times (\bra{\phi_0}_{c_1r_1}\bra{\phi_2^t}_{c_2r_2}
+\bra{\phi_1^t}_{c_1r_1}\bra{\phi_3^t}_{c_2r_2})],
\end{eqnarray}
where the components can be written as
\begin{eqnarray}\label{9}
\ket{\phi_0}&=&\ket{00},\nonumber\\
\ket{\phi_1^t}&=&\xi\ket{10}+\chi\ket{01},\nonumber\\
\ket{\phi_2^t}&=&\xi^2\ket{20}+\sqrt{2}\xi\chi\ket{11}+\chi^2\ket{02},\nonumber\\
\ket{\phi_3^t}&=&\xi^3\ket{30}+\sqrt{3}\xi^2\chi\ket{21}+\sqrt{3}\xi\chi^2\ket{12}+\chi^3\ket{03},
\end{eqnarray}
in which the amplitudes are $\xi(t)=e^{-\kappa t/2}$ and $\chi(t)=(1-e^{-\kappa t})^{1/2}$, with the
parameter $\kappa$ being the dissipative constant \cite{note1}.

In this section, we focus on the dynamic properties of two cavities which are initially in an MMES.
As the cavities and reservoirs interact, the state of two cavities is
$\rho_{c_1c_2}(t)=\mbox{Tr}_{r_1r_2}[\rho_{c_1r_1c_2r_2}(t)]$, which has the matrix form
\begin{equation}\label{10}
\rho_{c_1c_2}(t)=\left(
                   \begin{array}{cccccccc}
                     a_{11} & 0 & 0 & 0 & 0 & a_{16} & 0 & 0 \\
                     0 & a_{22} & 0 & 0 & 0 & 0 & a_{27} & 0 \\
                     0 & 0 & a_{33} & 0 & 0 & 0 & 0 & a_{38} \\
                     0 & 0 & 0 & a_{44} & 0 & 0 & 0 & 0 \\
                     0 & 0 & 0 & 0 & a_{55} & 0 & 0 & 0 \\
                     a_{61} & 0 & 0 & 0 & 0 & a_{66} & 0 & 0 \\
                     0 & a_{72} & 0 & 0 & 0 & 0 & a_{77} & 0 \\
                     0 & 0 & a_{83} & 0 & 0 & 0 & 0 & a_{88} \\
                   \end{array}
                 \right)
\end{equation}
with the basis in the order
$\{\ket{00},\ket{01},\ket{02},\ket{03},\ket{10},\ket{11},\ket{12},\ket{13}\}$ and the matrix elements
\begin{eqnarray}\label{11}
&&a_{11}=(p+\chi^4+\chi^8-p\chi^8)/2,\nonumber\\
&&a_{22}=\xi^2\chi^2[2-p+3(1-p)\chi^4]/2,\nonumber\\
&&a_{33}=(1-p)\xi^4(1+3\chi^4)/2,\nonumber\\
&&a_{44}=(1-p)\xi^6\chi^2/2,\nonumber\\
&&a_{55}=\xi^2\chi^2(p+\chi^4-p\chi^4)/2,\nonumber\\
&&a_{66}=\xi^4[p+3(1-p)\chi^4]/2,\nonumber\\
&&a_{77}=3(1-p)\xi^6\chi^2/2,\nonumber\\
&&a_{88}=(1-p)\xi^8/2,\nonumber\\
&&a_{16}=a_{61}=\xi^2[p+\sqrt{3}(1-p)\chi^4]/2,\nonumber\\
&&a_{27}=a_{72}=\sqrt{3/2}(1-p)\xi^4\chi^2,\nonumber\\
&&a_{38}=a_{83}=(1-p)\xi^6/2.
\end{eqnarray}

In order to characterize the dynamic entanglement properties of two cavities, we need to choose a
suitable measure of entanglement. Here, we use the negativity \cite{vidal02pra} to quantify the
entanglement of two cavities due to its computability for any state of an arbitrary bipartite
system. For the quantum state $\rho_{c_1c_2}(t)$, its negativity is
\begin{equation}\label{12}
N_{c_1c_2}(t)=\frac{||\rho^{T_{c_1}}_{c_1c_2}(t)||-1}{2}=\frac{\sum_{i=1}^{8}|\lambda_i|-1}{2},
\end{equation}
where $||\cdot||$ is the trace norm and equal to the sum of the moduli of the eigenvalues for the
Hermitian matrix $\rho^{T_{c_1}}_{c_1c_2}(t)$, which is the partial transpose with respect to the
subsystem $c_1$ \cite{vidal02pra}. After some derivation, we can obtain the eigenvalues
\begin{eqnarray}\label{13}
&&\lambda_1=(1-p)\xi^8/2,\nonumber\\
&&\lambda_2=(p+\chi^4+\chi^8-p\chi^8)/2,\nonumber\\
&&\lambda_3=\xi^4\{[1+(6-6p)\chi^4]-\sqrt{A}\}/4,\nonumber\\
&&\lambda_4=\xi^4\{[1+(6-6p)\chi^4]+\sqrt{A}\}/4,\nonumber\\
&&\lambda_5=[2(1-p)\xi^6\chi^2-\sqrt{B}]/2,\nonumber\\
&&\lambda_6=[2(1-p)\xi^6\chi^2+\sqrt{B}]/2,\nonumber\\
&&\lambda_7=\xi^2\{\chi^2[1+2(1-p)\chi^4]-\sqrt{C}\}/2,\nonumber\\
&&\lambda_8=\xi^2\{\chi^2[1+2(1-p)\chi^4]+\sqrt{C}\}/2,
\end{eqnarray}
where the parameters are $A=(1-2p)^2+24(1-p)^{2}\chi^4$, $B=(1-p)\xi^{12}(1+\chi^4)$, and
$C=p^2+(1-p)[1+(2\sqrt{3}-1)p]\chi^4+5(1-p)^2\chi^2+(1-p)^2\chi^{12}$.

\begin{figure}
\epsfig{figure=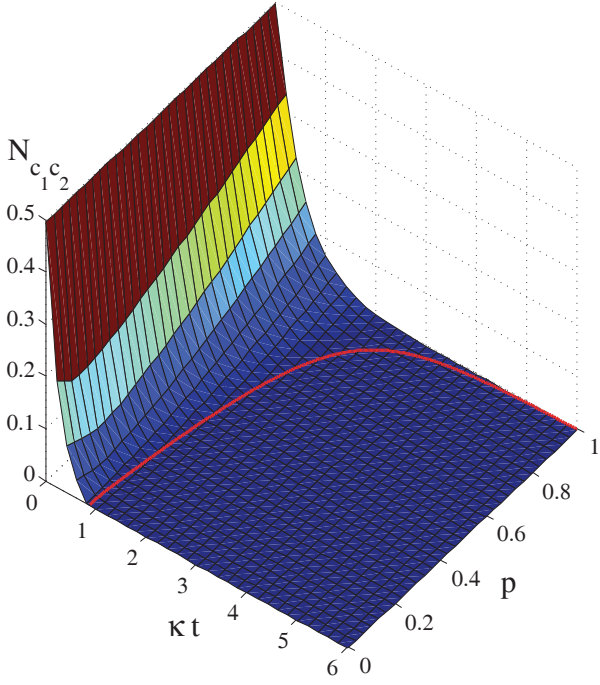,width=0.45\textwidth}
\caption{(Color online) Evolution of entanglement of the MMES in two cavities, where the negativity
is shown as a function of the time $\kappa t$ and the probability $p$. The red line indicates the 
entanglement sudden death of the two cavities.}
\end{figure}

In Fig. 1, we plot the negativity $N_{c_1c_2}(t)$ as a function of the time $\kappa t$ and
the probability $p$. As seen from Fig. 1, when $\kappa t=0$, the quantum state
$\rho_{c_1c_2}$ is the MMES, and its negativity has the maximal value of 0.5 regardless of the choice of
probability $p$. For a given value of the probability, $p$, the negativity decreases as
the time, $\kappa t$, increases. It should be pointed out that as time increases, the entanglement
of the MMES decays through sudden death rather than asymptotically like the two-qubit Bell state.
The red line in Fig.1 indicates the time of the ESD for the negativity $N_{c_1c_2}(t)$ and satisfies
the equation
\begin{equation}\label{14}
p=\frac{1-\sqrt{3}+3\chi^4-3\chi^8+\sqrt{D}}
{1-2\sqrt{3}+\chi^{-4}+5\chi^4-3\chi^8}
\end{equation}
where the parameter $D=3-2\sqrt{3}+4(2-\sqrt{3})\chi^4+\chi^8$, with
$\chi=\sqrt{1-e^{-\kappa t}}$ (the derivation of the ESD line is presented in Appendix A). When the
probability $p=0$, the initial state of the two cavities is $\ket{\psi_2}=(\ket{02}+\ket{13})/\sqrt{2}$,
which is a two-qubit pure maximal entangled state with qubit $c_1$ involving states $0$ and $1$  and
qubit $c_2$ involving states $2$ and $3$.
As the system evolves, the quantum state of two cavities becomes a $2\otimes4$ system and has the
matrix form shown in Eq.(10) with the parameter $p=0$. Its entanglement evolution shows the ESD
phenomenon, and the negativity  becomes zero at the time
$\kappa t=\mbox{ln}[(3+\sqrt{3})/2]\approx0.8612$. When the probability $p\in(0,1)$, the
initial state $\rho_{c_1c_2}(0)$ is the MMES. The ESD time of the two cavities is determined by Eq. (14)
and increases as a function of the probability $p$. In the $p=1$ case, the initial state is
the two-qubit Bell state $\ket{\psi_1}=(\ket{00}+\ket{11})/\sqrt{2}$, and its entanglement
disappears at the time $\kappa t\rightarrow \infty$, which coincides with the result for asymptotic
decay presented by L\'{o}pez \emph{et al.} \cite{lopez08prl}.

Here, we have shown that, unlike the asymptotic entanglement decay of the Bell state, the MMES of
two cavities experiences the ESD in the dissipative procedure of cavity-reservoir systems.
It is argued that the high-dimensional component $\ket{\psi_2}=\ket{02}+\ket{13})/\sqrt{2}$ plays
the dominant role. Although the initial state $\ket{\psi_2}$ is a logic two-qubit state, it will
evolve to a $2\otimes 4$ system along with the cavity-reservoir interaction, which results in the
ESD phenomenon of two cavities. In the evolution of two cavities, the ESD time is postponed when
the mixed probability of the component $\ket{\psi_1}$ (the Bell state) increases. In the case of
$p=1$ for the MMES, the component $\ket{\psi_2}$ disappears, and there is no ESD phenomenon,
which is just the evolution of the pure Bell state $\ket{\psi_1}=(\ket{00}+\ket{11})/\sqrt{2}$.

In addition to quantum entanglement, nonlocality is also a useful resource in quantum secure
communication. It is worthwhile to further investigate whether maximally entangled states like
the MMES also result in maximal nonlocality and how the nonlocality of the MMES evolves with time. 
The MIN \cite{sluo11prl} is a computable nonlocality measure, which is the maximum global effect 
caused by locally invariant measurement. The MIN is defined as
\cite{sluo11prl}
\begin{equation}\label{15}
\mbox{MIN}(\rho_{AB})=\mbox{max}_{\Pi^A}||\rho-\Pi^A(\rho_{AB})||^2,
\end{equation}
where the max runs over all the von Neumann measurements $\Pi^A=\{\Pi_k^A\}$ which do not disturb
the reduced density matrix $\rho_A$ (\emph{i.e.}, $\sum_k\Pi_k^A\rho_A\Pi_k^A=\rho_A$), and the
Hilbert-Schmidt norm is $||X||^2=\mbox{tr}X^{\dagger}X$.
The state of the two cavities $\rho_{c_1c_2}(t)$ in Eq. (10) can be
rewritten in a generalized Bloch form,
\begin{eqnarray}\label{16}
\rho_{c_1c_2}(t)&=&\frac{1}{2\sqrt{2}}\frac{I_2}{\sqrt{2}}\otimes \frac{I_4}{2}+\sum_{i=1}^{3}x_iX_i
\otimes\frac{I_4}{2}+\frac{I_2}{\sqrt{2}}\nonumber\\&&
\otimes\sum_{j=1}^{15}y_jY_j+\sum_{i=1}^{3}\sum_{j=1}^{15} T_{ij}X_{i}\otimes Y_{j},
\end{eqnarray}
where $I_2$ and $I_4$ are identity matrices of the subsystems and $X_i=\sigma_i/\sqrt{2}$ and
$Y_j=(\sigma_{j'}\otimes\sigma_{j''})/2$ are operator bases with $j',j''\in\{0,1,2,3\}$ except
for the case $j'=j''=0$ (here, $\sigma_0=I_2$ and $\{\sigma_1,\sigma_2,\sigma_3\}$ are the Pauli
matrices). In the Bloch expression, Eq. (16), the matrix elements are
\begin{eqnarray}\label{17}
x_i&=&\mbox{tr}(\rho_{c_1c_2}X_i\otimes I_4/2),\nonumber\\
y_j&=&\mbox{tr}(\rho_{c_1c_2}I_2/\sqrt{2}\otimes Y_j),\nonumber\\
T_{ij}&=&\mbox{tr}(\rho_{c_1c_2}X_i\otimes Y_j).
\end{eqnarray}

Luo and Fu derived an analytical formula for the MIN in an arbitrary $2\otimes d$
system \cite{sluo11prl}
\begin{numcases}
{\mbox{MIN}_{c_1c_2}=}
{\mbox{tr}TT^t-\frac{1}{\parallel \textbf{x}\parallel^2} \textbf{x}^tTT^t\textbf{x}}&
if $\textbf{x}\neq 0$ \nonumber\\
{\mbox{tr}TT^t-\lambda_3}& \textrm{if $\textbf{x}=0$}
\end{numcases}
where $\lambda_3$ is the minimum eigenvalue of the $3\times 3$ matrix $TT^t$ with
$T=(T_{ij})$, and $\textbf{x}=(x_1,x_2,x_3)^t$ is the local Bloch vector with the norm
$||\textbf{x}||^2=\sum_ix_i^2$ (here, $t$ represents the transposition). After some derivation, we can
obtain the expression for the MIN of two dissipative cavities, which can be written as
\begin{equation}\label{19}
\mbox{MIN}_{c_1c_2}(t)=\frac{1}{2}\xi^4[F-2p(F-G)+p^2(1-2G+F)],
\end{equation}
where the two parameters are $F=\xi^8+6\xi^4\chi^4+3\chi^8$ and $G=\sqrt{3}\chi^4$ with
$\xi=e^{-\kappa t/2}$ and $\chi=\sqrt{1-e^{-\kappa t}}$. The details of the calculation and
the continuity analysis of the MIN are presented in Appendix B.

\begin{figure}
\epsfig{figure=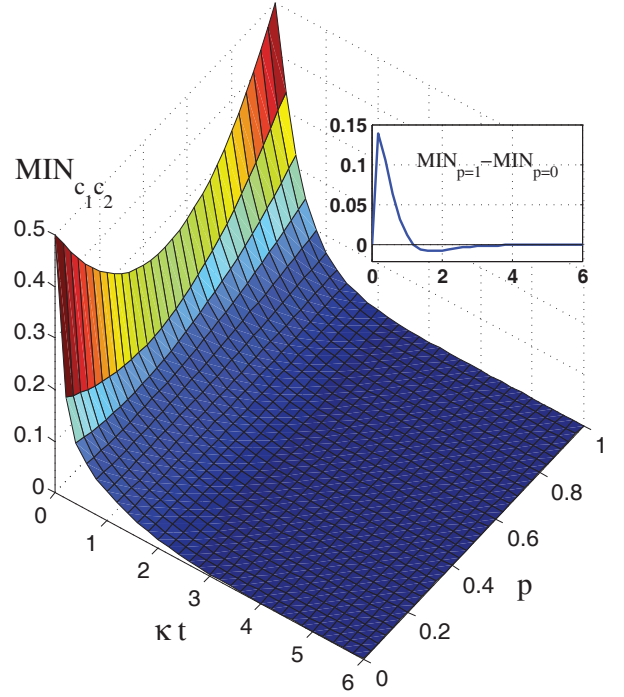,width=0.45\textwidth}
\caption{(Color online) Evolution of the MIN of the MMES in two cavities, where the nonlocality is
shown as a function of the time $\kappa t$ and the probability $p$. The inset is the difference
$\mbox{MIN}_{p=1}-\mbox{MIN}_{p=0}$ as a function of $\kappa t$.}
\end{figure}

In Fig.2, we plot the MIN as a function of the time $\kappa t$ and the probability $p$.
When $\kappa t=0$, $\rho_{c_1c_2}(0)$ is the MMES, and its nonlocality is
\begin{equation}\label{20}
\mbox{MIN}_{c_1c_2}(0)=\left(p-\frac{1}{2}\right)^2+\frac{1}{4},
\end{equation}
which is symmetric to the probability $p=1/2$.
As shown in Fig. 2, the MIN has the maximum value of $0.5$ for the cases of $p=0$ and
$p=1$, which correspond to the pure maximally entangled states $\ket{\psi_2}$ and $\ket{\psi_1}$.
When $p\in (0,1)$, the MIN has less than the maximum value, and reaches the minimum value of 
$0.25$ at $p=0.5$. Therefore,
we can reach the conclusion that the nonlocality of the MMES is not maximal, although its
entanglement is maximal for any value of the probability $p$.
According to Eq. (20), we find that the MIN is directly proportional to the purity
$\mbox{Tr}(\rho^2)$ of the MMES,
\emph{i.e.}, $\mbox{MIN}_{c_1c_2}(0)=\frac{1}{2}\mbox{Tr}[\rho_{c_1c_2}^2(0)]$. When
the mixed-state probability of the MMES changes from $0$ to $1/2$,  its purity decreases, which
results in the MIN changing from the maximum of $0.5$ to the minimum of $0.25$. When the probability $p$
changes from $1/2$ to $1$, the purity of the MMES also increases, and the MIN changes from the
minimum of $0.25$ to the maximum of $0.5$. As two cavities evolve, the $\mbox{MIN}_{c_1c_2}(t)$ decays
in an asymptotic manner and disappears in the
limit $\kappa t\rightarrow \infty$. This is different from the sudden-death evolution of the
negativity of two cavities, since the nonlocality described by the MIN contains both quantum and
classical correlations. The inset of Fig. 2 shows the difference $\mbox{MIN}_{p=1}-\mbox{MIN}_{p=0}$,
which indicates that the nonlocality is no longer symmetric as the system evolves.

\section{Entanglement and nonlocality distributions of the MMES in multipartite dynamics}

Entanglement monogamy is an important property of multipartite systems and means that quantum
entanglement cannot be freely shared among many parties
\cite{ben96pra,ckw00pra,osb06prl,byw07pra,bxw14prl,jens15prl}. It has been proved that the
squared negativity obeys the monogamy inequality in pure states of qubit systems,
$N_{A_1|A_2\cdots A_n}^2-N_{A_1A_2}^2-\cdots-N_{A_1A_n}^2\geq0$ \cite{oufan07pra}. However, for
mixed states or multilevel pure-state systems, whether or not a similar monogamy relation holds
is still an open problem. Recently, a numerical analysis was carried out for tripartite multilevel 
pure states \cite{hevidal15pra}, which supported the monogamy relations for squared negativity.
However, it is still unknown
whether or not the monogamous relation holds for the four-partite case, especially in a real quantum
system with dissipative reservoirs. With this in mind, we next analyze the negativity distribution
of the MMES in the four-partite $2\otimes 2\otimes 4\otimes 4$ cavity-reservoir systems.
On the one hand, this analysis can verify the monogamy inequality for the squared negativity, and,
on the other hand, it can provide a deep understanding of the dynamics of the MMES.

The residual entanglement in monogamy inequalities can be used as a multipartite
entanglement measure or indicator to characterize the structure of multipartite entanglement
\cite{lohmayer06prl,baw08pra,byw08pra2,reg14prl,arg14prl}. For composite cavity-reservoir systems,
we analyze the entanglement distribution in the partition $c_1r_1|c_2r_2$ and evaluate the
multipartite entanglement indicator
\begin{eqnarray}\label{21}
M_{c_1r_1|c_2r_2}(t)&=&N_{c_1r_1|c_2r_2}^2(t)-N_{c_1c_2}^2(t)-N_{c_1r_2}^2(t)\nonumber\\
&&-N_{r_1c_2}^2(t)-N_{r_1r_2}^2(t).
\end{eqnarray}
As the system evolves, the four-partite negativity is invariant and satisfies the relation
$N_{c_1r_1|c_2r_2}(t)=N_{c_1r_1|c_2r_2}(0)=N_{c_1c_2}(0)=0.5$ since the local dissipation is unitary
and the two reservoirs are in the vacuum state initially. At a later time, the state of the two
reservoirs has a form similar to that of the two cavities, and we get the relation
$\rho_{r_1r_2}(t)=S_{\xi\leftrightarrow\chi}[\rho_{c_1c_2}(t)]$ in which $S_{\xi\leftrightarrow\chi}$
is an operation exchanging two parameters (\emph{i.e.}, $\xi\rightarrow\chi$ and $\chi\rightarrow\xi$).
Thus the negativity of the reservoirs is
\begin{equation}\label{22}
N_{r_1r_2}(t)=S_{\xi\leftrightarrow\chi}[N_{c_1c_2}(t)].
\end{equation}
We can also derive the relationship
$\rho_{r_1c_2}(t)=S_{\xi\leftrightarrow\chi}[\rho_{c_1r_2}(t)]$ and the negativities
\begin{equation}\label{23}
N_{r_1c_2}(t)=S_{\xi\leftrightarrow\chi}[N_{c_1r_2}(t)]
\end{equation}
for subsystems $c_1r_2$ and $r_1c_2$. A more detailed description of the density matrix
$\rho_{c_1r_2}(t)$ and its negativity $N_{c_1r_2}(t)$ can be found in Appendix C.

\begin{figure}
\epsfig{figure=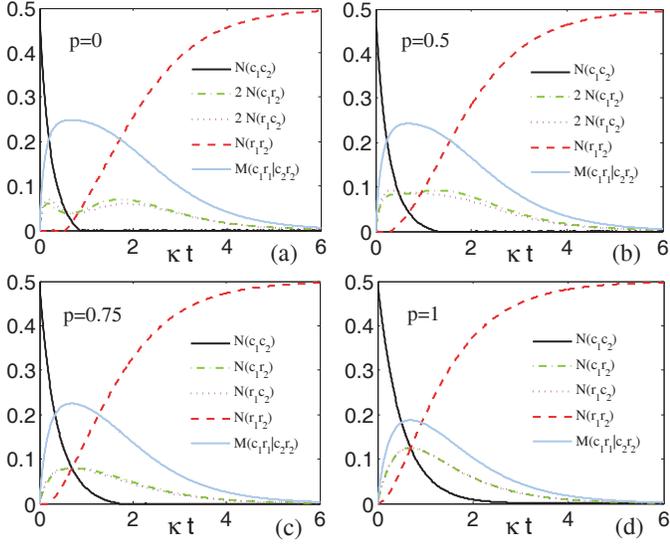,width=0.5\textwidth}
\caption{(Color online) The distribution of negativity in the dissipation of multipartite $2\otimes
2\otimes 4 \otimes 4$ cavity-reservoir systems, where the non-negative values of $M_{c_1r_1|c_2r_2}$
indicate multipartite entanglement.}
\end{figure}

In Fig. 3, we plot the negativity distribution as a function of the time $\kappa t$ for the cases
where the probability $p$ of the MMES is chosen to be $0, 0.5, 0.75$, and $1$. As time increases,
the two reservoirs exhibit the phenomenon of entanglement sudden birth (ESB) \cite{lopez08prl},
which corresponds to the ESD of the two cavities. As the probability $p$ increases, the ESB time is
advanced, and the ESD time is delayed, as shown in Figs. 3(a)-3(c). When the probability is $p=1$, both
the ESB and ESD phenomena disappear, as shown in Fig. 3(d), since the initial state becomes the
two-qubit Bell state. For the subsystems $c_1r_2$ and $r_1c_2$, the negativities have two peak values
when the probability is $p=0$ and $p=0.5$ [see Figs. 3(a) and 3(b), where we multiply $N_{c_1r_2}$ and
$N_{r_1c_2}$ by a factor of 2 for clarity]. As the probability $p$ increases, the number of peaks in
the negativity changes from two to one, as shown in Figs. 3(c) and 3(d). We further calculated the 
entanglement distribution in Eq. (21) and found that the squared negativity is monogamous
in the composite cavity-reservoir systems. Therefore, the quantity $M_{c_1r_1|c_2r_2}(t)$ can serve
as a multipartite entanglement indicator as time evolves, as plotted (solid-blue line) in Fig. 3 for
different probabilities.

Next, we analyze the MIN distribution of the MMES in the multipartite cavity-reservoir system. It has
been proved that the MIN is not monogamous in multiqubit systems \cite{asen12jpa,qlin15cpb}. However,
whether the MIN is monogamous in multipartite multilevel systems needs to be further
investigated, especially for the newly introduced MMES. Using the relationships of the density
matrices $\rho_{c_1c_2}$, $\rho_{c_1r_2}$, $\rho_{r_1c_2}$, and $\rho_{r_1r_2}$, we can get
\begin{eqnarray}\label{24}
&&\mbox{MIN}_{r_1r_2}(t)=S_{\xi\leftrightarrow\chi}[\mbox{MIN}_{c_1c_2}(t)],\nonumber\\
&&\mbox{MIN}_{r_1c_2}(t)=S_{\xi\leftrightarrow\chi}[\mbox{MIN}_{c_1r_2}(t)],
\end{eqnarray}
where $S_{\xi\leftrightarrow\chi}$ is the exchanging operation acting on the parameters $\xi$ and
$\chi$. After a derivation similar to that for $\mbox{MIN}_{c_1c_2}(t)$, we can obtain the MIN of
the subsystem $c_1r_2$,
\begin{eqnarray}\label{25}
\mbox{MIN}_{c_1r_2}(t)=\frac{1}{2}\xi^2\chi^2[p^2+2\sqrt{3}p(1-p)\xi^4+F_1],
\end{eqnarray}
with the parameter $F_1=(3\xi^8+6\xi^4\chi^4+\chi^8)(1-p)^2$. In addition, for the MIN of multipartite
cavity-reservoir systems in the partition $c_1r_1|c_2r_2$, we can obtain the expression
\begin{equation}\label{26}
\mbox{MIN}_{c_1r_1|c_2r_2}(t)=\mbox{MIN}_{c_1r_1|c_2r_2}(0)=\frac{1}{2}(1-2p+2p^2),
\end{equation}
where the MIN is invariant as the time increases because the evolution operation $U_{c_1r_1}(\hat{H},t)
\otimes U_{c_2r_2}(\hat{H},t)$ is locally unitary. In addition, we calculate the MIN distribution
$M'_{c_1r_1|c_2r_2}(t)=\mbox{MIN}_{c_1r_1|c_2r_2}(t)-\mbox{MIN}_{c_1c_2}(t)-\mbox{MIN}_{c_1r_2}(t)
-\mbox{MIN}_{r_1c_2}(t)-\mbox{MIN}_{r_1r_2}(t)$, which is written as
\begin{eqnarray}\label{27}
M'_{c_1r_1|c_2r_2}(t)=(1-p)\xi^2\chi^2(G_1-\sqrt{3}p),
\end{eqnarray}
with the parameter $G_1=(1-p)(1-\chi^2+\chi^4)$.

\begin{figure}
\epsfig{figure=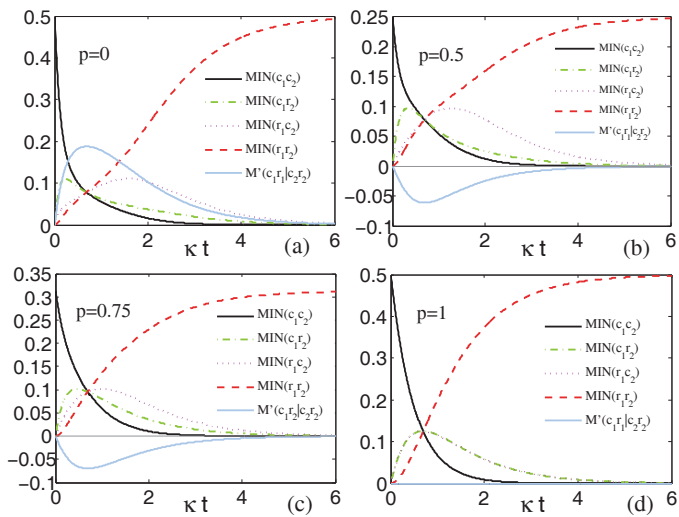,width=0.5\textwidth}
\caption{(Color online) The MIN distribution of the MMES in multipartite $2\otimes 2\otimes 4 \otimes 4$
cavity-reservoir systems, where the residual nonlocality $M'_{c_1r_1|c_2r_2}$ can be positive, zero,
or negative as a function of the time parameter $\kappa t$.}
\end{figure}

In Fig. 4, we plot the distribution of the MIN as a function of time $\kappa t$ with different
probabilities $p$ for the MMES. As time increases, the MIN of two cavities decreases asymptotically,
and the MIN of the two reservoirs increases asymptotically. When the time $\kappa t\rightarrow \infty$,
the MIN of the two cavities transfers completely to the reservoirs. For the subsystems $c_1r_2$ and
$r_1c_2$, the MINs first increase to their maximums and then decay asymptotically. As the probability
increases, the distance between the two peaks of $\mbox{MIN}_{c_1r_2}$ and $\mbox{MIN}_{r_1c_2}$ becomes
smaller. When the probability is $p=1$, the distance goes to zero, and the two MINs coincide
completely, as shown in Fig. 4(d). Unlike the distribution of entanglement negativity, the MIN in the
multipartite systems is not always monogamous. When the mixed-state probability is $p=0$, the MIN is
monogamous and the indicator $M'_{c_1r_1|c_2r_2}(t)$ (the solid blue line) is nonnegative, as
shown in Fig. 4(a). However, when the probabilities are $p=0.5$ and $p=0.75$, the MINs are polygamous,
and $M'_{c_1r_1|c_2r_2}(t)$ is no longer positive [see Fig. 4(b) and 4(c)]. When the probability
is $p=1$, the MMES becomes the two-qubit Bell state, and the indicator $M'_{c_1r_1|c_2r_2}(t)$ is
zero at all times for cavity-reservoir systems as shown in Fig. 4(d). The different distribution 
property from that of entanglement negativity indicates that the MIN and entanglement are two 
inequivalent types of resources for quantum information processing.

Although the MIN itself is not monogamous in multipartite multilevel systems, its functions may
possess this property. For example, the quantum discord \cite{oll01prl,ved01jpa} is not monogamous
even in three-qubit pure states \cite{prabhu12pra,stre12prl,gio11pra,fan13pra}, but the squared
quantum discord is monogamous in an arbitrary three-qubit pure state \cite{bzyw13pra,sli14ann}.
Recently, similar situations for the entanglement of formation have also been discussed
\cite{bxw14prl,oli14pra,xzhu14pra,bxw14pra,yluo15arxiv,song16pra}.
For the MIN in multipartite cavity-reservoir
systems, we calculated the square of the MIN, and the numerical result supports the monogamy relation.
However, in the general case, an analytical proof for the monogamy property of the squared MIN
is still an open problem.

\section{discussion and conclusion}

We have studied the dynamic behavior of the MMES over the course of the dissipative
evolution of multipartite multilevel cavity-reservoir systems. It has been found that, unlike the
two-qubit Bell state $\ket{\psi_1}=(\ket{00}+\ket{11})/\sqrt{2}$, whose negativity decays in an
asymptotic manner \cite{lopez08prl}, the entanglement dynamics of the MMES exhibits the ESD phenomenon,
as shown in Fig.  1. Therefore, as an entanglement resource, the MMES is not superior
to the pure two-qubit Bell state in this dissipative system. We think that the high-dimensional
component $\ket{\psi_2}=(\ket{02}+\ket{13})/\sqrt{2}$ in the MMES gives rise to the ESD of the two 
cavities' evolution. Moreover, we further study the MMESs in $2\otimes 6$ and $2\otimes 8$ systems 
where the component $\ket{\psi_2}$ is replaced by the higher-dimensional components
$\ket{\psi_3}=(\ket{04}+\ket{15})/\sqrt{2}$ and $\ket{\psi_4}=(\ket{06}+\ket{17})/\sqrt{2}$,
respectively.
The analytical results show that the new MMESs still experience the ESD in the dynamical
evolution (the details for the calculation are given in Appendix D), which further supports
our viewpoint.

The MIN has a close relation with quantum communication protocols involving local measurement and
a comparison between the pre- and postmeasurement states \cite{sluo11prl}. We find that maximal
entanglement \emph{cannot} guarantee maximal nonlocality. As shown in Fig. 2, the MIN of the MMES
is not maximal, and its value is directly proportional to the purity $\mbox{Tr}(\rho^2)$ of the MMES,
which is quite different from the situation of the Bell state exhibiting maximal nonlocality.
For the MMESs with higher-dimensional components, their MINs are also dependent on the mixed state
probability $p$, and the nonlocality evolutions are asymptotical (the details are given in
Appendix E). We explain that the decrease of the MIN of the MMESs is due to the decrease of their
purity, and the MIN evolution of the MMESs is asymptotic since this kind of nonlocality contains both
quantum and classical correlations \cite{sluo11prl}. For the quantum nonlocality related
to the violation of Bell inequalities \cite{gho09prl,aug10prl,gall12prl,vert12prl,bus12prl,liang14prl},
its relation to the maximal entangled state is still an open problem yet to be addressed.

In order to obtain a deep understanding of the dynamic properties of the MMES, we have investigated
its entanglement and nonlocality distributions in multipartite systems. The numerical results have
shown that the squared negativity is monogamous in multipartite cavity-reservoir systems
(beyond the four values of probability $p$ shown in Fig.3, we further calculated the distribution
for $p$ ranging across $[0,1]$). Moreover, for the MMESs of $2\otimes 6$ and $2\otimes 8$ systems,
our numerical calculation still shows that the squared negativity is monogamous in the multipartite
dissipative systems. These results support the conjecture of He and Vidal \cite{hevidal15pra} that
the squared negativity is monogamous in multipartite multilevel systems.
On the other hand, the MIN distribution of the MMES is not monogamous in the multipartite
cavity-reservoir system, as shown in Fig. 4, which indicates that the MIN is a different type of
resource from entanglement in quantum information processing. We further investigate the MIN
distributions for the MMESs with higher-dimensional components and find that the MIN is still not
monogamous (the details are shown in Appendix E).

In conclusion, we have studied the dynamic behavior of the MMES in multipartite cavity-reservoir
systems. It has been found that the evolution of the negativity of the MMES exhibits the ESD
phenomenon, and is not superior to the two-qubit Bell state as an entanglement resource in a
dissipative system. We also found that maximal entanglement cannot guarantee maximal nonlocality.
The MIN of the MMES is not maximal, and its evolution is dependent on the mixed-state probability
of the MMES. In addition, we have investigated the distributions of the negativity and
the MIN of the MMESs in the multipartite cavity-reservoir systems, where two types of correlation
exhibit different monogamous properties.

\section*{Acknowledgments}

Y.-K. Bai is grateful to Prof. N. Davison for critical reading of the manuscript and also
would like to thank Y.-F. Xu for many useful discussions. This work was supported by the URC fund 
of HKU, NSF-China under Grants No. 11575051 and No. 11274124, Hebei NSF under Grant No. A2016205215, 
and the funds of Hebei Normal University under Grants No. L2008B03 and No. L2012B06.

\appendix

\section{The derivation of the ESD line for negativity $N_{c_1c_2}$}

In Eq. (12) of the main text, the negativity $N_{c_1c_2}(t)$ is determined via the sum of absolute
values of the negative eigenvalues. After some analysis, we find that the eigenvalues
$\{\lambda_1,\lambda_2,\lambda_4,\lambda_6,\lambda_8\}$ are always nonnegative, while the other
eigenvalues $\{\lambda_3,\lambda_5,\lambda_7\}$ can be positive, zero, or negative. Therefore,
as the two cavities evolve, the negativity $N_{c_1c_2}(t)$ becomes zero when the three eigenvalues
$\{\lambda_3,\lambda_5,\lambda_7\}$ become nonnegative.

Using the expressions for $\lambda_3$, $\lambda_5$, and $\lambda_7$ in Eq. (13), we can derive
the $p\sim\kappa t$ relations when these eigenvalues are zero. When $\lambda_3=0$, we have the relation
\begin{equation}\label{A1}
p=\frac{3(e^{\kappa t}-1)^2(3-6e^{\kappa t}+2e^{2\kappa t})}{9-36e^{\kappa t}+48e^{2\kappa
t}-24e^{3\kappa t}+2e^{4\kappa t}}.
\end{equation}
For the case $\lambda_5=0$, we have
\begin{equation}\label{A2}
\kappa t=\mbox{ln}[(3+\sqrt{3})/2]
\end{equation}
for an arbitrary value of parameter $p$.
When $\lambda_7=0$, we can derive the $p\sim\kappa t$ relation as shown in Eq. (14) of the main text.
In Fig.5, we plot the three relations in the plane of parameters $p$ and $\kappa t$, where the dashed
green line is for $\lambda_3=0$, the dot-dashed blue line is for $\lambda_5=0$, and the solid red line
is for $\lambda_7=0$. The three lines divide the whole area into four parts. In regions I, II and III,
the signs of the eigenvalues $(\lambda_3,\lambda_5,\lambda_7)$ are $(-,-,-)$, $(+,-,-)$, and $(+,+,-)$,
which result in nonzero negativity for the two cavities. In region IV, all the eigenvalues are positive
leading to the negativity being $N_{c_1c_2}(t)=0$. Thus, as seen from Fig. 5, the red line for
$\lambda_7=0$ determines the ESD time of the two cavities, which is described by Eq. (14) of
the main text.

\begin{figure}
\epsfig{figure=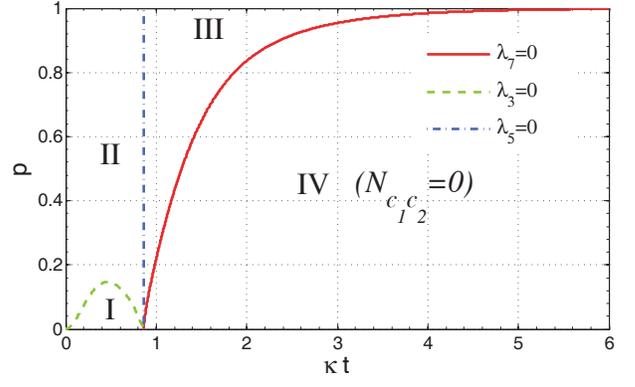,width=0.45\textwidth}
\caption{(Color online) Four regions in the entanglement evolution of two cavities, where, in region
IV, all three eigenvalues $(\lambda_3,\lambda_5,\lambda_7)$ are positive and the negativity
$N_{c_1c_2}$ becomes zero.}
\end{figure}

\section{Calculation and continuity analysis for the MIN of two cavities}

Before evaluating the nonlocality $\mbox{MIN}_{c_1c_2}(t)$ given in Eq. (18), we first calculate the
local
Bloch vector $\textbf{x}$ and correlation matrix $T$. According to Eq. (17), the local Bloch vector
of subsystem $c_1$ is
\begin{equation}\label{B1}
\textbf{x}=\left(0,0,\frac{\chi^2}{2\sqrt{2}}\right)^t,
\end{equation}
which leads to the norm being $||\textbf{x}||^2=\chi^4/8$, with $\chi=\sqrt{1-e^{-\kappa t}}$. The
correlation matrix $T=T'/2\sqrt{2}$ is a $3\times 15$ matrix, in which the nonzero elements
of $T'$ are
\begin{eqnarray}\label{B2}
T'_{1,1}&=&-T_{2,2}=\xi^2[p+(1-p)\xi^4+\sqrt{3}(1-p)\chi^4],\nonumber\\
T'_{3,3}&=&(1-2\chi^2+2\chi^4)[1-4(1-p)\xi^2\chi^2],\nonumber\\
T'_{1,5}&=&T'_{1,10}=T'_{3,6}=-T'_{2,9}=\sqrt{6}(1-p)\xi^4\chi^2,\nonumber\\
T'_{1,13}&=&\xi^2[p-(1-p)\xi^4+\sqrt{3}(1-p)\chi^4],\nonumber\\
T'_{2,14}&=&-\xi^2[p-(1-p)\xi^4+\sqrt{3}(1-p)\chi^4],\nonumber\\
T'_{3,12}&=&\chi^2-4(1-p)\xi^2\chi^4,\nonumber\\
T'_{3,15}&=&2p-1+(6p-4)\chi^2-(8-10p)\chi^4\nonumber\\
&&+6(1-p)\chi^6,
\end{eqnarray}
with $\xi=e^{-\kappa t/2}$.
When $\kappa t>0$, we have the local Bloch vector $\textbf{x}\neq0$. After substituting the three
terms $\textbf{x}$, $||\textbf{x}||^2$, and $T$ into the first formula in Eq. (18), we can obtain the
expression for $\mbox{MIN}_{c_1c_2}(t>0)$ in Eq. (19). When $\kappa t=0$, the quantum state
$\rho_{c_1c_2}(0)$ is the MMES for which the local Bloch vector is $\textbf{x}=0$. In this case, we
need to calculate the eigenvalues of matrix $TT^t$, which are
$\lambda_1=\lambda_2=\lambda_3=(1-2p+2p^2)/4$. According to the second formula in Eq. (18), we can
derive $\mbox{MIN}_{c_1c_2}(0)=(1-2p+2p^2)/2$.

Next, we prove the continuity of $\mbox{MIN}_{c_1c_2}(t)$. Based on the previous analysis, we know that
the two formulas in Eq. (18) are used for the cases $\textbf{x}\neq 0$ and  $\textbf{x}=0$, which
correspond to the time evolutions $\kappa t>0$ and $\kappa t=0$, respectively. If the
$\mbox{MIN}_{c_1c_2}$ is continuous, the limit of $\mbox{MIN}_{c_1c_2}(\kappa t\rightarrow 0_+)$ in
the first formula should coincide with the value of $\mbox{MIN}_{c_1c_2}(\kappa t=0)$. After some
calculation, we can get
\begin{eqnarray}\label{B3}
\lim_{\kappa t\rightarrow 0_+} \frac{\textbf{x}^tTT^t\textbf{x}}{\parallel \textbf{x}\parallel^2}
&=&\lim_{\kappa t\rightarrow 0_+} \frac{\frac{d(\textbf{x}^tTT^t\textbf{x})}{d(\kappa t)}}
{\frac{d(\parallel \textbf{x}\parallel^2)}{d(\kappa t)}}\nonumber\\
&=&\lim_{\kappa t\rightarrow 0_+} \frac{\frac{d^2(\textbf{x}^tTT^t\textbf{x})}{d(\kappa t)^2}}
{\frac{d^2(\parallel \textbf{x}\parallel^2)}{d(\kappa t)^2}}\nonumber\\
&=&\frac{\frac{1}{16}(1-2p+2p^2)}{\frac{1}{4}}\nonumber\\
&=&\lambda_3,
\end{eqnarray}
where we have used L'H\^{o}pital's rule and $\lambda_3$ is the minimal eigenvalue of matrix $TT^t$.
Then, in the limit $\kappa t\rightarrow 0_+$, the two formulas in Eq. (18) are continuous and we have
\begin{eqnarray}\label{B4}
\lim_{\kappa t\rightarrow 0_+}\mbox{MIN}_{c_1c_2}(\kappa t)&=&\mbox{MIN}_{c_1c_2}(0)\nonumber\\
&=&\frac{1}{2}(1-2p+2p^2).
\end{eqnarray}
As a result, the nonlocality $\mbox{MIN}_{c_1c_2}(t)$ can be described by Eq. (19) of the main text
throughout the entire period of the dynamic evolution.

\section{The density matrix $\rho_{c_1r_2}$ and its negativity $N_{c_1r_2}$}
Throughout the dynamic evolution of multipartite cavity-reservoir systems, the quantum state of subsystem
$c_1r_2$ is $\rho_{c_1r_2}(t)=\mbox{tr}_{r_1c_2}[\rho_{c_1r_1c_2r_2}(t)]$, which can be written as
\begin{equation}\label{C1}
\rho_{c_1r_2}(t)=\left(
                   \begin{array}{cccccccc}
                     b_{11} & 0 & 0 & 0 & 0 & b_{16} & 0 & 0 \\
                     0 & b_{22} & 0 & 0 & 0 & 0 & b_{27} & 0 \\
                     0 & 0 & b_{33} & 0 & 0 & 0 & 0 & b_{38} \\
                     0 & 0 & 0 & b_{44} & 0 & 0 & 0 & 0 \\
                     0 & 0 & 0 & 0 & b_{55} & 0 & 0 & 0 \\
                     b_{61} & 0 & 0 & 0 & 0 & b_{66} & 0 & 0 \\
                     0 & b_{72} & 0 & 0 & 0 & 0 & b_{77} & 0 \\
                     0 & 0 & b_{83} & 0 & 0 & 0 & 0 & b_{88} \\
                   \end{array}
                 \right),
\end{equation}
where the nonzero matrix elements are
\begin{eqnarray}\label{C2}
&&b_{11}=[p+(1-p)\xi^4](1+\xi^2\chi^2)/2,\nonumber\\
&&b_{22}=\{2(1-p)\xi^2\chi^2+[p+3(1-p)\xi^4]\chi^4\}/2,\nonumber\\
&&b_{33}=(1-p)\chi^4(1+3\xi^2\chi^2)/2,\nonumber\\
&&b_{44}=(1-p)\chi^8/2,\nonumber\\
&&b_{55}=[\xi^8+p(\xi^4-\xi^8)]/2,\nonumber\\
&&b_{66}=\xi^2\chi^2[p+3(1-p)\xi^4]/2,\nonumber\\
&&b_{77}=3(1-p)\xi^4\chi^4/2,\nonumber\\
&&b_{88}=(1-p)\xi^2\chi^6/2,\nonumber\\
&&b_{16}=b_{61}=\xi\chi[p+\sqrt{3}(1-p)\xi^4]/2,\nonumber\\
&&b_{27}=b_{72}=\sqrt{3/2}(1-p)\xi^3\chi^3,\nonumber\\
&&b_{38}=b_{83}=(1-p)\xi\chi^5/2.
\end{eqnarray}
For this quantum state, the negativity is
\begin{equation}\label{C3}
N_{c_1r_2}(t)=\frac{\sum_{i=1}^8|\lambda_i|-1}{2}
\end{equation}
where $\lambda_i$ are the eigenvalues of the partial transpose matrix $\rho_{c_1r_2}^{T_{c_1}}$
and have the form
\begin{eqnarray}\label{C4}
&&\lambda_1=(1-p)\xi^2\chi^6/2,\nonumber\\
&&\lambda_2=(1+\xi^2\chi^2)[p+(1-p)\xi^4]/2,\nonumber\\
&&\lambda_3=[1-2\chi^2+B_1-\sqrt{1-(1-p)\chi^2B_2}]/4,\nonumber\\
&&\lambda_4=[1-2\chi^2+B_1+\sqrt{1-(1-p)\chi^2B_2}]/4,\nonumber\\
&&\lambda_5=[(3-2p)\chi^2+B_3-\sqrt{\chi^4B_4}]/4,\nonumber\\
&&\lambda_6=[(3-2p)\chi^2+B_3+\sqrt{\chi^4B_4}]/4,\nonumber\\
&&\lambda_7=[(1-p)\chi^4(3\xi^4+\chi^4)-\sqrt{(1-p)^2B_5}]/4,\nonumber\\
&&\lambda_8=[(1-p)\chi^4(3\xi^4+\chi^4)+\sqrt{(1-p)^2B_5}]/4,
\end{eqnarray}
with the parameters
\begin{eqnarray}\label{C5}
B_1&=&(7-5p)\chi^4-10(1-p)\chi^6+(4-4p)\chi^8,\nonumber\\
B_2&=&(16-8\sqrt{3})p+[14-24(2-\sqrt{3})p]\chi^2\nonumber\\
&&-8[8-(10-3\sqrt{3})p]\chi^4+[123-(111-8\sqrt{3})p]\chi^6\nonumber\\
&&-(104-96p)\chi^8+28(1-p)\chi^{10}+(8-8p)\chi^{12}\nonumber\\
&&-4(1-p)\chi^{14},\nonumber\\
B_3&=&-(8-7p)\chi^4+12(1-p)\chi^6-6(1-p)\chi^8,\nonumber\\
B_4&=&(3-2p)^2-[36-2(23-6p)p]\chi^2\nonumber\\
&&+(64-96p+33p^2)\chi^4-12(1-p)(4-3p)\chi^6\nonumber\\
&&-12(1-p)^2\chi^8,\nonumber\\
B_5&=&\chi^8(9\xi^8+4\xi^2\chi^2-6\xi^4\chi^4+\chi^8).
\end{eqnarray}

\section{The ESD for the MMESs with higher-dimensional components}

In the multipartite cavity-reservoir systems, we first consider that the two cavities are initially in
the MMES
\begin{equation}\label{D1}
\rho_{c_1c_2}^{(1)}(0)=p\proj{\psi_1}+(1-p)\proj{\psi_3},
\end{equation}
where $\ket{\psi_1}=(\ket{00}+\ket{11})/\sqrt{2}$ is the two-qubit Bell state and
$\ket{\psi_3}=(\ket{04}+\ket{15})/\sqrt{2}$ is the high dimensional component.
Along with the evolution of cavity-reservoir systems, the output state is
\begin{eqnarray}\label{D2}
\rho_{c_1r_1c_2r_2}^{(1)}(t)
&=&\frac{p}{2}[(\ket{\phi_0}_{c_1r_1}\ket{\phi_0}_{c_2r_2}
+\ket{\phi_1^t}_{c_1r_1}\ket{\phi_1^t}_{c_2r_2})\nonumber\\
&&\times(\bra{\phi_0}_{c_1r_1}\bra{\phi_0}_{c_2r_2}
+\bra{\phi_1^t}_{c_1r_1}\bra{\phi_1^t}_{c_2r_2})]\nonumber\\
&+&\frac{1-p}{2}[(\ket{\phi_0}_{c_1r_1}\ket{\phi_4^t}_{c_2r_2}
+\ket{\phi_1^t}_{c_1r_1}\ket{\phi_5^t}_{c_2r_2})\nonumber\\
&&\times(\bra{\phi_0}_{c_1r_1}\bra{\phi_4^t}_{c_2r_2}
+\bra{\phi_1^t}_{c_1r_1}\bra{\phi_5^t}_{c_2r_2})],
\end{eqnarray}
where the components have the forms
\begin{eqnarray}\label{D3}
\ket{\phi_0^t}&=&\ket{00},\nonumber\\
\ket{\phi_1^t}&=&\xi\ket{10}+\chi\ket{01},\nonumber\\
\ket{\phi_4^t}&=&\xi^4\ket{40}+2\xi^3\chi\ket{31}+\sqrt{6}\xi^2\chi^2\ket{22}\nonumber\\
              &&+2\xi\chi^3\ket{13}+\chi^4\ket{04},\nonumber\\
\ket{\phi_5^t}&=&\xi^5\ket{50}+\sqrt{5}\xi^4\chi\ket{41}+\sqrt{10}\xi^3\chi^2\ket{32}\nonumber\\
              &&\sqrt{10}\xi^2\chi^3\ket{23}+\sqrt{5}\xi\chi^4\ket{14}+\chi^5\ket{05},
\end{eqnarray}
with the parameters $\xi(t)=e^{-\kappa t/2}$ and $\chi(t)=(1-e^{-\kappa t})^{1/2}$.
By tracing the subsystems of two reservoirs, we can get the output state of two cavities
$\rho_{c_1c_2}^{(1)}(t)=\mbox{Tr}_{r_1r_2}[\rho_{c_1r_1c_2r_2}^{(1)}(t)]$, which is a $12\times 12$
matrix. In order to obtain the entanglement negativity of $\rho_{c_1c_2}^{(1)}(t)$, we calculate the
eigenvalues of the partial transpose matrix $\rho_{c_1c_2}^{(1)T_{c_1}}(t)$. After some derivation,
we find that there are four eigenvalues which can be negative,
\begin{eqnarray}\label{D4}
\lambda_2&=&(1-p)\xi^{10}(3\chi^2-\sqrt{1+4\chi^4})/2,\nonumber\\
\lambda_6&=&(1-p)\xi^8(1+15\chi^4-\sqrt{1+70\chi^4+25\chi^8})/4,\nonumber\\
\lambda_7&=&(1-p)\xi^6\chi^2(1+5\chi^4-\sqrt{1+15\chi^4}),\nonumber\\
\lambda_{11}&=&\xi^2\chi^2[\chi^4(2+3\chi^4)+p(1-2\chi^4-3\chi^8)]/2\nonumber\\
           &&-\sqrt{H_1}/2,
\end{eqnarray}
where the parameter is
$H_1=\xi^4[p^2+2\sqrt{5}(1-p)p\chi^8+(1-p)^2(4\chi^{12}+13\chi^{16}+4\chi^{20})]$.
Similar to the analysis in Appendix A, we can derive the ESD time for the MMES $\rho_{c_1c_2}^{(1)}(t)$
according to the four eigenvalues. When the mixed-state probability $p$ changes in the region
$[0,p_1]$ with $p_1=(347-125\sqrt{5})/1922\approx 0.03512$, the ESD time for the MMES is
$\kappa t=\mbox{ln}[(5+\sqrt{5})/4]\approx 0.5928$. When the probability $p\in[p_1,1)$, the ESD time
is determined by the $p\sim\kappa t$ relation
\begin{eqnarray}\label{D5}
p=\frac{\sqrt{2}\chi^8\sqrt{J_1}+\chi^8J_2}{1-\chi^4+2(2-\sqrt{5})\chi^8+6\chi^{12}+\chi^{16}-5\chi^{20}}
\end{eqnarray}
where the two parameters are $J_1=4-2\sqrt{5}+3(3-\sqrt{5})\chi^4+2\chi^8$ and
$J_2=2-\sqrt{5}+3\chi^4+\chi^8-5\chi^{12}$. In Fig.6(a), we plot the ESD line (red line) as a function
$p(\kappa t)$, which divides the entanglement evolution into an entangled region and a disentangled region.

\begin{figure}
\epsfig{figure=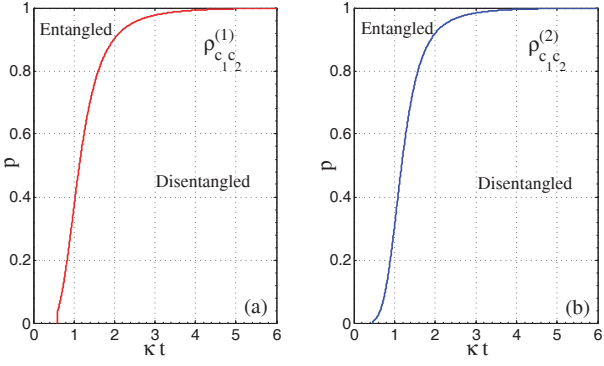,width=0.45\textwidth}
\caption{(Color online) The ESD lines in the evolution of two cavities which are initially in
the MMESs: (a) $\rho_{c_1c_2}^{(1)}$ in Eq. (D1) and (b) $\rho_{c_1c_2}^{(2)}$ in Eq. (D6).}
\end{figure}

Next, we consider the two cavities which are initially in the MMES,
\begin{equation}\label{D6}
\rho_{c_1c_2}^{(2)}(0)=p\proj{\psi_1}+(1-p)\proj{\psi_4},
\end{equation}
where $\ket{\psi_4}=(\ket{06}+\ket{17})/\sqrt{2}$ is the high-dimensional component.
As the systems evolves, the output state $\rho_{c_1r_1c_2r_2}^{(2)}(t)$ has the same form as that
in Eq. (D2), but the components $\ket{\phi_4^t}$ and $\ket{\phi_5^t}$ are replaced by the new
components $\ket{\phi_6^t}$ and $\ket{\phi_7^t}$, which can be written as
\begin{eqnarray}\label{D7}
\ket{\phi_6^t}&=&\xi^6\ket{60}+\sqrt{6}\xi^5\chi\ket{51}+\sqrt{15}\xi^4\chi^2\ket{42}\nonumber\\
                 &&+\sqrt{20}\xi^3\chi^3+\sqrt{15}\xi^2\chi^4+\sqrt{6}\xi\chi^5\ket{15}\nonumber\\
                 &&+\chi^6\ket{06},\nonumber\\
\ket{\phi_7^t}&=&\xi^7\ket{70}+\sqrt{7}\xi^6\chi\ket{61}+\sqrt{21}\xi^5\chi^2\ket{52}\nonumber\\
      &&+\sqrt{35}\xi^4\chi^3\ket{43}+\sqrt{35}\xi^3\chi^4\ket{34}+\sqrt{21}\xi^2\chi^5\ket{25}\nonumber\\
      &&+\sqrt{7}\xi\chi^6\ket{16}+\chi^7\ket{07}.
\end{eqnarray}
After tracing the subsystems $r_1r_2$, we can get the output state of two cavities
$\rho_{c_1c_2}^{(2)}(t)$. Furthermore, by doing the partial transposition, we can obtain the matrix
$\rho_{c_1c_2}^{(2)T_{c_1}}(t)$ and calculate its eigenvalues. The ESD line is determined by the
negative eigenvalues of $\rho_{c_1c_2}^{(2)T_{c_1}}(t)$. When the mixed state probability
$p\in[0,p_2]$, with $p_2=(8669-2401\sqrt{7})/370191\approx 0.006258$, the ESD occurs at the time
$\kappa t=\mbox{ln}[(7+\sqrt{7})/6]\approx 0.4748$. When $p\in[p_2,1)$, the ESD time is determined by
the following $p\sim \kappa t$ relation:
\begin{equation}\label{D8}
p=\frac{\chi^{12}(K_1+\sqrt{K_2})}{1-\chi^4+2(3-\sqrt{7})\chi^{12}+8\chi^{16}+\chi^{24}-7\chi^{28}}
\end{equation}
where the two parameters are $K_1=3-\sqrt{7}+4\chi^4+\chi^{12}-7\chi^{16}$ and
$K_2=15-6\sqrt{7}+8(4-\sqrt{7})\chi^4+9\chi^8$. In Fig.6(b), we plot the ESD line (blue line) as
a function $p(\kappa t)$, which cut the entanglement evolution region into two parts,
\emph{i.e.}, an entangled region and a disentangled region.

\section{The MIN of the MMES with higher-dimensional components and its distribution}

We first consider the MMES $\rho_{c_1c_2}^{(1)}(0)$ with the high-dimensional component
$\ket{\psi_3}=(\ket{04}+\ket{15})/\sqrt{2}$, as shown in Eq. (D1). According to the formula in Eq. (18)
of the main text, we can derive the nonlocality
\begin{equation}\label{E1}
\mbox{MIN}[\rho_{c_1c_2}^{(1)}(0)]=(p-1/2)^2+1/4,
\end{equation}
which is dependent on the mixed-state probability $p$ and directly proportional to the purity of the
MMES. In the calculation of the MIN, the matrix basis for the subsystem $c_2$ is chosen to be the
generalized Gell-Mann matrices (GGM) \cite{bert08jpa}, which are the higher-dimensional extension of
the Pauli matrices. The GGM basis for a $d$-dimensional system is composed of three types of
matrices \cite{bert08jpa}:
(i) $d(d-1)/2$ symmetric GGM,
\begin{eqnarray}\label{E2}
\Lambda_{s}^{jk}=\ket{j}\bra{k}+\ket{k}\bra{j}, ~~~~1\leq j<k\leq d;
\end{eqnarray}
(ii) $d(d-1)/2$ antisymmetric GGM,
\begin{eqnarray}\label{E3}
\Lambda_{a}^{jk}=-i\ket{j}\bra{k}+i\ket{k}\bra{j}, ~~~~1\leq j<k\leq d;
\end{eqnarray}
and (iii) $(d-1)$ diagonal GGM,
\begin{eqnarray}\label{E4}
\Lambda^{l}=\sqrt{2/(l^2-l)}(\sum_{j=1}^{l}\proj{j}-l\proj{l+1})
\end{eqnarray}
with $1\leq l \leq d-1$. It should be noted that the GGM needs to be normalized in the generalized
Bloch form of $\rho_{c_1c_2}^{(1)}$.
Along with the interaction between the cavities and reservoirs, the MMES will evolve
into a $2\otimes 6$ system. After some derivation, we obtain
\begin{equation}\label{E5}
\mbox{MIN}[\rho_{c_1c_2}^{(1)}(t)]=\frac{\xi^4}{2}\{\xi^{16}+p[p-(2-p)\xi^{16}]+L_1+L_2\}
\end{equation}
where the two parameters are $L_1=2(1-p)[\sqrt{5}p+30(1-p)\xi^8]\chi^8$ and
$L_2=(1-p)^2(20\xi^{12}\chi^4+40\xi^4\chi^{12}+5\chi^{16})$. In the dissipative procedure of
cavity-reservoir systems, the nonlocality of two cavities decays in an asymptotical way, which is
similar to the case of MMES in $2\otimes 4$ systems.

\begin{figure}[t]
\epsfig{figure=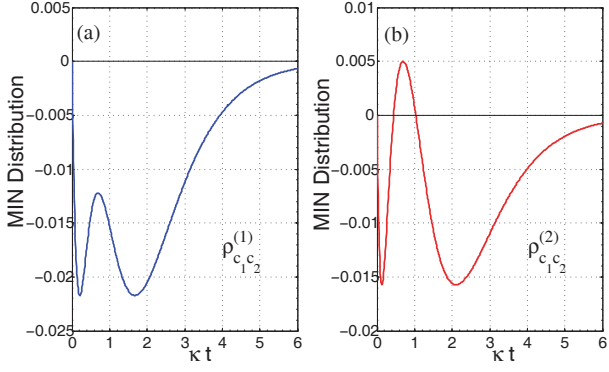,width=0.45\textwidth}
\caption{(Color online) The MIN distributions of the MMESs in multipartite
$2\otimes 2\otimes 6\otimes 6$ and $2\otimes2\otimes 8\otimes 8$ cavity-reservoir systems
are plotted as a function of $\kappa t$, where the negative values indicate that the MIN
is not monogamous.}
\end{figure}

Next, we analyze the MMES of a $2\otimes 8$ system in Eq. (D6), which has the high-dimensional
component $\ket{\psi_4}=(\ket{06}+\ket{17})/\sqrt{2}$. It is found that the MIN for this MMES
$\rho_{c_1c_2}^{(2)}(0)$ has the same
expression as that in Eq. (E1), which is also dependent on the mixed-state probability $p$.
As the system evolves, the MIN for two cavities decays asymptotically and can be expressed as
\begin{eqnarray}\label{E6}
\mbox{MIN}[\rho_{c_1c_2}^{(2)}(t)]=\frac{\xi^4}{2}[\xi^{24}+L_3+(1-p)^2L_4],
\end{eqnarray}
where the parameters are $L_3=p[p-(2-p)\xi^{24}]+2(1-p)[\sqrt{7}p+350(1-p)\xi^{12}]\chi^{12}$ and
$L_4=42\xi^{20}\chi^4+315\xi^{16}\chi^8+525\xi^8\chi^{16}+126\xi^4\chi^{20}+7\chi^{24}$.

For the MMESs $\rho_{c_1c_2}^{(1)}(0)$ and $\rho_{c_1c_2}^{(2)}(0)$ with the high-dimensional
components, we further calculate the distribution
\begin{eqnarray}\label{E7}
M'_{c_1r_1|c_2r_2}(t)&=&\mbox{MIN}_{c_1r_1|c_2r_2}(t)-\mbox{MIN}_{c_1c_2}(t)\nonumber\\
                      &&-\mbox{MIN}_{c_1r_2}(t)-\mbox{MIN}_{r_1c_2}(t)\nonumber\\
                      &&-\mbox{MIN}_{r_1r_2}(t)
\end{eqnarray}
in the multipartite cavity-reservoir systems. We find that the MIN
distributions are still not monogamous. As examples, we choose the mixed-state probability $p=0.8$
for the two MMESs and calculate their MIN distributions.
In Fig. 7, the MIN distributions in the multipartite systems are plotted , where two cavities are
initially in the MMESs $\rho_{c_1c_2}^{(1)}(0)$  and $\rho_{c_1c_2}^{(2)}(0)$. As shown,
the negative values for the distributions indicate that the MIN is not monogamous.

However, for the squared negativity of the MMESs in $2\otimes 6$ and $2\otimes 8$ systems, we calculate
the entanglement distribution in the multipartite $2\otimes 2\otimes 6\otimes 6$ and
$2\otimes 2\otimes 8\otimes 8$ cavity-reservoir systems, where the mixed-state
probability $p$ ranges across $[0,1]$. The numerical results still support that the negativity is
monogamous.


\begin{thebibliography}{99}

\bibitem{horo09rmp}      R. Horodecki, P. Horodecki, M. Horodecki, and K. Horodecki, Rev. Mod. Phys.
                         \textbf{81}, 865 (2009).

\bibitem{elts14jpa}      C. Eltschka and J. Siewert, J. Phys. A \textbf{47}, 424005 (2014).

\bibitem{cui12nc}        J. Cui, M. Gu, L. C. Kwek, M. F. Santos, H. Fan, and V. Vedral, Nature Commun.
                         \textbf{3}, 812 (2012).

\bibitem{ben93prl}       C. H. Bennett, G. Brassard, C. Cr\'{e}peau, R. Jozsa, A. Peres, and W. K.
                         Wootters, Phys. Rev. Lett. \textbf{70}, 1895 (1993).

\bibitem{ekert91prl}     A. K. Ekert, Phys. Rev. Lett. \textbf{67}, 661 (1991).

\bibitem{ben92prl}       C. H. Bennett and S. J. Wiesner, Phys. Rev. Lett. \textbf{69}, 2881 (1992).

\bibitem{cava05pra}      D. Cavalcanti, F. G. S. L. Brand\~{a}o, and M. O. Terra Cunha, Phys. Rev. A
                         \textbf{72}, 040303 (2005).

\bibitem{zgli12qic}      Z. G. Li, M. J. Zhao, S. M. Fei, H. Fan, and W. M. Liu, Quantum Inf. Comp.
                         \textbf{12}, 0063 (2012).

\bibitem{mjzhao15pra}    M.-J. Zhao, Phys. Rev. A \textbf{91}, 012310 (2015).

\bibitem{zycz01pra}      K. \.{Z}yczkowski, P. Horodecki, M. Horodecki, and R. Horodecki, Phys. Rev. A
                         \textbf{65}, 012101 (2001).

\bibitem{sche03jmo}      S. Scheel, J. Eisert, P. L. Knight, and M. B. Plenio, J. Mod.Opt. \textbf{50},
                         881 (2003).

\bibitem{tyu04prl}       T. Yu and J. H. Eberly, Phys. Rev. Lett. \textbf{93}, 140404 (2004).

\bibitem{alm07sci}       M. P. Almeida, F. de Melo, M. Hor-Meyll, A. Salles, S. P. Walborn, P. H. Souto
                         Ribeiro, and L. Davidovich, Science \textbf{316}, 579 (2007).

\bibitem{lau07prl}       J. Laurat, K. S. Choi, H. Deng, C. W. Chou, and H. J. Kimble, Phys. Rev. Lett.
                         \textbf{99}, 180504 (2007).

\bibitem{tyu09sci}       T. Yu and J. H. Eberly, Science \textbf{323}, 598 (2009).

\bibitem{lopez08prl}     C. E. L\'{o}pez, G. Romero, F. Lastra, E. Solano, and J. C. Retamal, Phys. Rev.
                         Lett. \textbf{101}, 080503 (2008).

\bibitem{ben99pra}       C. H. Bennett, D. P. DiVincenzo, C. A. Fuchs, T. Mor, E. Rains, P. W. Shor,
                         J. A. Smolin, and W. K. Wootters, Phys. Rev. A \textbf{59}, 1070 (1999).

\bibitem{buhr10rmp}      H. Buhrman, R.Cleve, S. Massar, and R. de Wolf, Rev. Mod. Phys. \textbf{82},
                         665, (2010).
\bibitem{sluo11prl}      S. Luo and S. Fu, Phys. Rev. Lett. \textbf{106}, 120401 (2011).

\bibitem{bell64phy}      J. S. Bell, Physics (N. Y.) \textbf{1}, 195 (1964).

\bibitem{clau69prl}      J. F. Clauser, M. A. Horne, A. Shimony, and R. A. Holt, Phys. Rev. Lett.
                         \textbf{23}, 880 (1969).

\bibitem{vidal02pra}     G. Vidal and R. F. Werner, Phys. Rev. A \textbf{65}, 032314 (2002).

\bibitem{byw09pra}       Y.-K. Bai, M.-Y. Ye, and Z. D. Wang, Phys. Rev. A \textbf{80}, 044301 (2009).

\bibitem{wen11epjd}      W. Wen, Y.-K. Bai, and H. Fan, Eur. Phys. J. D \textbf{64}, 557 (2011).

\bibitem{ali14pra}       M. Ali and A. R. P. Rau, Phys. Rev. A \textbf{90}, 042330 (2014).

\bibitem{note1}          In order to conveniently obtain the output state in Eq. (8), we can introduce
                         an environment system $E$ which purifies the quantum state $\rho_{c_1c_2}$. 
                         Then the global initial state can be written as $\ket{\Psi(0)}=(\sqrt{p}\ket{\psi_1}_{c_1c_2}\ket{0}_{E}+ \sqrt{1-q} \ket{\psi_2}_{c_1c_2}\ket{1}_E)\ket{00}_{r_1r_2}$, and the global output state
                         $\ket{\Psi(t)}$ can be obtained via the components in Eq. (9). After tracing
                         the environment system $E$, we can get the output state.

\bibitem{ben96pra}       C. H. Bennett, H. J. Bernstein, S. Popescu, and B. Schumacher, Phys. Rev. A.
                         \textbf{53}, 2046 (1996).

\bibitem{ckw00pra}       V. Coffman, J. Kundu, and W. K. Wootters, Phys. Rev. A \textbf{61}, 052306
                         (2000).

\bibitem{osb06prl}       T. J. Osborne and F. Verstraete, Phys. Rev. Lett. \textbf{96}, 220503 (2006).

\bibitem{byw07pra}       Y.-K. Bai, D. Yang, and Z. D. Wang, Phys. Rev. A \textbf{76}, 022336 (2007).

\bibitem{bxw14prl}       Y.-K. Bai, Y.-F. Xu, and Z. D. Wang, Phys. Rev. Lett. \textbf{113},
                         100503 (2014).

\bibitem{jens15prl}      C. Eltschka and J. Siewert, Phys. Rev. Lett. \textbf{114}, 140402 (2015).

\bibitem{oufan07pra}     Y.-C. Ou and H. Fan, Phys. Rev. A \textbf{75}, 062308 (2007).

\bibitem{hevidal15pra}   H. He and G. Vidal, Phys. Rev. A \textbf{91}, 012339 (2015).

\bibitem{lohmayer06prl}  R. Lohmayer, A. Osterloh, J. Siewert, and A. Uhlmann, Phys. Rev. Lett.
                         \textbf{97}, 260502 (2006).

\bibitem{baw08pra}       Y.-K. Bai and Z. D. Wang, Phys. Rev. A \textbf{77}, 032313 (2008).

\bibitem{byw08pra2}      Y.-K. Bai, M.-Y. Ye, and Z. D. Wang, Phys. Rev. A \textbf{78}, 062325 (2008).

\bibitem{reg14prl}       B. Regula, S. Di Martino, S. Lee, and G. Adesso, Phys. Rev. Lett. \textbf{113},
                         110501 (2014)

\bibitem{arg14prl}       G. H. Aguilar, A. Vald\'{e}s-Hern\'{a}ndez, L. Davidovich, S. P. Walborn, and
                         P. H. Souto Ribeiro, Phys. Rev. Lett. \textbf{113}, 240501 (2014).

\bibitem{asen12jpa}      A. Sen, D. Sarkar, and A. Bhar, J. Phys. A \textbf{45}, 405306 (2012).

\bibitem{qlin15cpb}      Q. Lin, Y.-K. Bai, M.-Y. Ye, and X.-M. Lin, Chin. Phys. B \textbf{24},
                         030304 (2015).

\bibitem{oll01prl}       H. Ollivier and W. H. Zurek, Phys. Rev. Lett. \textbf{88}, 017901 (2001).

\bibitem{ved01jpa}       L. Henderson and V. Vedral, J. Phys. A \textbf{34}, 6899 (2001).

\bibitem{prabhu12pra}    R. Prabhu, A. K. Pati, A. Sen(De), and U. Sen, Phys. Rev. A \textbf{85},
                         040102 (2012).

\bibitem{stre12prl}      A. Streltsov, G. Adesso,M. Piani, and D. Bru\ss, Phys. Rev. Lett. \textbf{109},
                         050503 (2012).

\bibitem{gio11pra}       G. L. Giorgi, Phys. Rev. A \textbf{84}, 054301 (2011).

\bibitem{fan13pra}       F. F. Fanchini, M. C. de Oliveira, L. K. Castelano, and M. F. Cornelio, Phys.
                         Rev. A \textbf{87}, 032317 (2013).

\bibitem{bzyw13pra}      Y.-K. Bai, N. Zhang, M.-Y. Ye, and Z. D. Wang, Phys. Rev. A \textbf{88},
                         012123 (2013).

\bibitem{sli14ann}       K. Salini, R. Prabhu, A. Sen(De), and U. Sen, Ann. Phys. (N.Y.) \textbf{348},
                         297 (2014).

\bibitem{oli14pra}       T. R. de Oliveira, M. F. Cornelio, and F. F. Fanchini, Phys. Rev. A
                         \textbf{89}, 034303 (2014).

\bibitem{xzhu14pra}      X.-N. Zhu and S.-M. Fei, Phys. Rev. A \textbf{90}, 024304 (2014).

\bibitem{bxw14pra}       Y.-K. Bai, Y.-F. Xu, and Z. D. Wang, Phys. Rev. A \textbf{90}, 062343 (2014).

\bibitem{yluo15arxiv}    Y. Luo and Y. Li, Ann. Phys. (N.Y.) \textbf{362}, 511 (2015).

\bibitem{song16pra}      W. Song, Y.-K. Bai, M. Yang, M. Yang, and Z.-L. Cao, Phys. Rev. A \textbf{93},
                         022306 (2016).

\bibitem{gho09prl}       S. Ghose, N. Sinclair, S. Debnath, P. Rungta, and R. Stock, Phys. Rev. Lett.
                         \textbf{102}, 250404 (2009).

\bibitem{aug10prl}       R. Augusiak, D. Cavalcanti, G. Prettico, and A. Ac\'{i}n, Phys. Rev. Lett.
                         \textbf{104}, 230401 (2010).

\bibitem{gall12prl}      R. Gallego, L. E. W\"{u}rflinger, A. Ac\'{i}n, and M. Navascu\'{e}s, Phys. Rev.
                         Lett. \textbf{109}, 070401 (2012).

\bibitem{vert12prl}      T. V\'{e}rtesi and N. Brunner, Phys. Rev. Lett. \textbf{108}, 030403 (2012).

\bibitem{bus12prl}       F. Buscemi, Phys. Rev. Lett. \textbf{108}, 200401 (2012).

\bibitem{liang14prl}     Y.-C. Liang, F. J. Curchod, J. Bowles, and N. Gisin, Phys. Rev. Lett.
                         \textbf{113}, 130401 (2014).

\bibitem{bert08jpa}      R. A. Bertlmann and P. Krammer, J. Phys. A \textbf{41},
                         235303 (2008).








\end{thebibliography}
\end{document}